# Identifying Nanotechnology in Society[1]

Charles Tahan
*Cavendish Laboratory, University of Cambridge,
JJ Thomson Ave, Cambridge, CB3 0HE, UK*
charlie@tahan.com
(2006)

**Abstract**

Manufacturing materials and systems with components thousands of times smaller than the width of a human hair promises vast and sometimes unimaginable advances in technology. Yet the term nanotechnology has formed as much from people's expectations as from scientific reality. Understanding the creation and context of this social construction can help us appreciate and guide what may be a burgeoning revolution. This chapter considers what different groups are referring to when they say nanotechnology, how this relates to the science involved, and how the various definitions of this broad field of endeavor might be improved. The ramifications and implications of these seemingly innocuous naming choices are also discussed. Although in many respects nanotechnology serves as cover justification for increased spending in the physical sciences, at present it is the most hopeful route to solving some of the planet's greatest problems.

**Table of Contents**

## 1. Introduction

When Norio Taniguchi first referred to his work as "nano-technology" (Taniguchi, 1974), it's doubtful he imagined that scholars might one day be pondering the word and considering its implications on humanity, even on the fate of humanity. But this year research on the societal implications of nanotechnology accounts for nearly 10% of direct federal funding on nanotechnology in the United States: 80% of that on environmental and toxicological effects and the remaining on broader sociological studies. (Roco, 2003b) That's not pocket change out of a budget of over $1 billion USD. But how can you study the societal impact of nanotechnology when the the term itself has not been clearly defined? In many ways, the label "nanotech" has been thrust on science and engineering externally, formed by society before solidifying as a field in its own right. This timing mismatch has interesting consequences for the shape of progress in nano-related fields and enterprises.

---

[1] Preprint of chapter to appear in *Advances in Computers*, Marvin Zelkowitz, ed. (2007).

A nanometer is one billionth of a meter or roughly equal to 10 hydrogen atoms laid side-by-side. Many systems, from biological to chemical to quantum, exhibit fundamentally new behavior at this length regime, often for completely coincidental reasons. For example, the blood-brain barrier often fails to filter particles of nanometer size, while being very selective about larger objects. The nanometer regime also happens to be where the quantum properties of electrons start to dominate in miniaturized circuits. The two have no connection in cause, but they are both considered "new nanoscale phenomena." This *reality of nanotechnology* contrasts with the *vision of nanotechnology* promoted in recent decades by futurists, science fiction writers, and doomsayers, as well as actual scientists. But *the reality* and *the vision* of nanotechnology have become entwined culturally despite their almost extreme differences.

Growing excitement in the field often traces back to two people: physicist Richard P. Feynman and researcher K. Eric Drexler. Physicists often justify their work by bowing to a past physics god, of which Feynman is one of the most famous. In his 1959 talk *There's Plenty of Room at the Bottom: An Invitation to Enter a New Field of Physics* (Feynman, 1959), Feynman laid out the vision of what would one day be called nanotechnology. Simultaneously simplistic and prescient, Feynman explored the vast technical applications that mastering matter on a small scale would create. He gave crude but surprisingly insightful suggestions about how to access and take advantage of all the room "at the bottom." Although one must be careful of fulfilling Matthew's Law (automatically attributing good ideas in a field to the most famous person in that field),[2] Feynman's talk anticipated manufacturing techniques such as the focused ion-beam etching Taniguchi was later to work on (and which is only now becoming mainstream in university labs), miniaturization in information storage, computation and machines, electron spin electronics (or spintronics), and even hinted at quantum computing.

Drexler is no physics god. But he deserves substantial credit for popularizing and motivating nanotechnology with his 1987 book *Engines of Creation: The Coming*

---

[2] The Gospel of Matthew (5-7) is where you find the recounting of the Sermon on the Mount. Matthew's Law is a bit of physics oral tradition; it refers to the tendency of the most well known person in a field to get credit for the best ideas in that field whether or not he had anything to do with it.

*Era of Nanotechnology* (Drexler, 1987) and later contributions. He falls squarely under the vision category of nanotechnology. The vision of nanotechnology could be described as atomic level systems engineering. Atoms, like so many Lego blocks, can be arranged one-by-one to build microscopic analogues of macroscopic machines or indeed anything one could imagine, from scratch. This is the life analog, the construction block view of both Feynman and Drexler, whom the latter calls machine-phase nanotechnology. From drugs designed from the bottom up to artificial nano-machines that could flow through and oxygenate blood on demand, the vision is dramatic. Drexler imagines assemblers, i.e. nanomachines that build other nanomachines, able to tackle problems from the size of the body to the globe. Take wood, an undisputedly useful building material "grown" by nanoscale/microscale machines. What else could be "grown" if we design the life that grows it? The implications, like those of biotechnology, can be scary. Fortunately, the vision is hard, and the reason this is so is because reality starts to intrude. The novel physics, surface science, and chemistry that begin to emerge in the nanoscale regime makes the construction block paradigms fall apart. The reality is actually much more interesting. Feynman himself pointed out that quantum effects and the complex behavior of many interacting particles – be they electrons, atoms, or molecules – would make this building block strategy difficult. In essence, that difficulty *is* nanotechnology today. But even he couldn't have imagined how big and how interesting those complications would be…and more importantly, how useful.

Let us pause for a brief moment to discuss bias and definitions. Many scientists could claim the authority to explain nanotechnology – biologists, chemists, materials scientists, and physicists. And all would necessarily do an incomplete job, a factor this chapter too will suffer from. Our bias is that of condensed matter and materials physics, the fields from which the transistor, solid-state laser, and silicon integrated circuit were born. Complex terms and definitions, unfortunately, will be commonplace in such a broad discussion of new technologies and their categorization.  Get used to being ignorant in this field. Brief definitions will be provided where possible (see the Appendix) but understand that scientists can use one word – say "spintronics" – to represent thousands of technical papers that may cover multiple sub-fields and have subtle meanings. Part of our challenge is that the forward line of the advancement of human knowledge has become so long. With so many individual innovators, complete categorization becomes impossible. You are encouraged to sleuth on your own when you encounter a cool sounding word you don't understand. Call it the joy of finding things out.

## 1.1     The scope of nanotechnology

There are many reports of varying quality that describe examples and applications of nanotechnology, from the perspectives of length-scale, medicine, environment, risk, society, and public policy. We especially recommend (SciAm,

2002) and (Amato, 2000). This chapter will describe the social construction of this term without becoming encyclopedic.

If the social construct of nanotechnology encompasses both the vision and the reality, one must consider the scope of both. Oddly, the vision of nanotechnology includes both the fantastic almost science-fiction visions of nano-assemblers and artificial life but also very non-nano visions including small machines and devices. Compared to a nanometer biological "machines" can be relatively huge; a red blood corpuscle is approximately 7000 nm while bacteria is 1000 nm and a virus is typically 60-100 nm. And key body dimensions are downright enormous. Swallowing a millimeter-sized pill that can take pictures of your insides, though it might be called nanotechnology, may have no nano-fabricated components whatsoever. This imprecision is present in both perspectives. In many ways the reality of what's going on in nanotechnology today – as presently named – is broader than even the vision.

At a recent conference on quantum nanoscience in Australia, a physics professor stood up and said something representative of many scientists' viewpoint: "Can't we just call it novel-technology?" Maybe we should. In the policy-technology world nanotechnology has become a catchall for all advanced materials technologies, from bio-integration to molecular transistors to quantum mesoscopics. The USA National Nanotechnology Initiative is an example of this, as its proponents seek to make nanotechnology a unifying theme across many of the sciences that deal with fabrication and manipulation of matter at small-scales.

Despite these facts-on-the-ground, some have claimed that atomic and molecular engineering alone is nanotechnology (Joachim, 2005). It's simply too late for this narrow and somewhat boring definition. Additionally, there are convincing reasons to lump larger systems under the nanotechnology roof. All the knowledge gained through the study of mesoscopic physics and complex systems might one day help the atomic-level construction technology. The science and technology community has to deal with putting these systems together and with the complex emergent behavior that implies. What quantum mesoscopics, for example, gives us is a physical window on nature's immense complication.

It would also be easy to say that the vision or molecular assembly view, as theorized by Drexler's so-called machine-phase nanotechnology, is the "true" nanotechnology and stop right here. But most of the money being spent on nanotechnology and most of the people who believe they are working on nanotechnology are in the reality camp. The glow that the term nanotechnology exudes not only attracts new students to science but also attracts scientists to adopt the term and become "nanotechnologists." And it helps with funding. Will nanotechnology just become a rallying cry or does it have any real meaning anymore?

From a societal standpoint, the consequences of this emerging collection of new artifacts, both directly as threats to the environment and our health, and indirectly via the transformative properties of life-changing technologies, have unquestionable importance. But there are limits to the extent of nanotechnology's embrace. In some situations, e.g. federal funding or patent decisions, more precise categorization can become necessary. For example the patent office makes a distinction that a new invention just can't be smaller to be patentable – there has to be something new there (like a fundamentally new manufacturing process). Many of the suggestions Feynman made just took advantage of making things smaller and were not necessarily based on new phenomena at that length scale. In the research world some inventions or lines of discovery, though they might coincidentally take place at the nanoscale, are better classified in a different scientific context. These are often one of disparate yet specific subfields. Specific examples will be presented in Section 3.

The reality of nanotechnology has matured at its own rate as technological manufacturing and measurement techniques have steadily improved over the last 50 years, putting fabrication and control of the nano realm within reach starting some 15 years ago (Eigler and Schweizer, 1990). It's easy to realize that "seeing" was the key to progress. Indeed, improvements in microscopes such as the electron microscope to the creation of new classes of devices such as atomic force microscopes (AFMs) and many other techniques contributed to advances in small science. Much of this is just the natural progress of science and technology, driven by human curiosity, economic interests like the semiconductor or drug industries, and luck.

The construction block view of nanotechnology is divorced from the daily interests of most scientists and so far away as to be utterly ignorable. In some sense it is a very simplistic idea. "Hey! Let's design things from the ground up!" Let's just make the nanoscopic equivalent of a macroscopic device. Replace the surgeon and scalpel with the mini-robot. It might be more difficult to do, for various reasons, but you are not necessarily using any new science to do it. Well, not so fast.

Complexity is an inherent problem in fields from computer science to biology to physics. It is *the* problem of the next one thousand years. Nanotechnology is not like building the first bridge. While it is clearly too soon for the vision of nanotechnology to be a wholly engineering endeavor, the beginnings are upon us. Nanoparticles like quantum dots and carbon nanotubes are the quintessential reality of nanotechnology today. They can exhibit new nanoscale phenomena in physics, chemical, or biological manifestations. Even if our theories of quantum physics, chemistry, and biology are complete, how they manifest themselves in complex situations is not. Progress in the nanoscience of mesoscopic and biological-materials systems will be the foundation of larger artificial small systems.

**Table 1: Nanotechnology circa 2006, vision versus reality, a pragmatic definition**

***The vision:*** From the fantastic to the mundane the vision of nanotechnology began over 40 years ago and encompasses the building block view of building matter and machines, and artificial life. Programmable machines you can't see.

*Examples:* atom-by-atom or molecule-by-molecule construction of everything from drugs to tiny robots; not necessarily nanometer-sized but also just small (e.g. cameras or surgical machines that can be swallowed); self-replicating nano-machines (or assemblers).

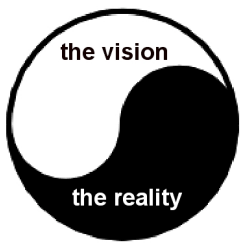

*Threats:* unknown, but greatly imagined: prolonging human life indefinitely; uncontrollable, replicating nano-assemblers turning the world into grey goo.

*Biggest questions:* Is this a joke? How will modifying or extending the human body change civilization? What about being able to "grow" anything cheaply with minimal materials?

***The reality:*** The science and technology behind nanotechnology today predominately defined as new phenomena that emerge in the nanometer length regime in physical, chemical, and biological systems.

*Examples:* nanoparticles that are 1) many times more reactive than microparticles, 2) have changing properties – like color – due to quantum confinement effects, or 3) have unknown biological effects like increased absorption in the body; advanced materials technology; quantum "mesoscopic" physics; everything and anything.

*Threats:* possible environmental hazard as nanoparticles can be much more reactive than larger particles and may linger longer; potential risk to human and animal health due to unknown effects on the body.

*Biggest questions:* How do we regulate these materials? How do we classify them and determine their risk?

***Nano-enabling technologies:*** Fabrication and measurement techniques that aid in the realization of the above two definitions but do not fall within their purview. The reliable manufacturing of devices and structures with at least two dimensions at the nanoscale.

## 1.2 Threats and futures and politics

Fear as much as hope has popularized nanotechnology, from the hysterical visions of "grey goo" to the quite realistic fears of human toxicity. The image of machines invisible to the naked eye wreaking havoc has certainly captured the public's imagination (Crichton, 2003; Joy, 2000). But aren't self-replicating robots that take over the world bad at any scale? In truth, many of the fears of robotics and artificial intelligence that have been around for decades have just been transferred to the invisible realm of nanotechnology, adding another dimension of fear. In any case, we already have replicating killer machines – they're called viruses, and bacteria, and humans. Redoing these with non-biological matter – almost the essence of the vision of nanotechnology – may well be more dangerous, or not, but will certainly be more difficult. We have to deal with those implications now in a plenty frightening way in biotechnology and genetic engineering. We're not sure what purpose calling what some biologists are already doing "nanotechnology" serves, except to say that it expands it to non-carbon based approaches.

Some of the new phenomena that nanoscale systems exhibit can be just as profoundly harmful as useful (Colvin, 2003; Hood, 2004). Particularly relevant are the chemical and biological effects. The same reactivity that makes nanoparticles powerful in military-grade explosives can also have dramatic impact on environmental systems. And nanotoxicity is an even more serious concern. Nanoparticle transmission and accumulation through the body is largely unknown. So investment into the study of environmental and health impact is necessary.

The real issue is why there has been so much funding for the societal implications of nanotechnology. The initial reason is to deal with the vision – but the fears of the vision are ridiculous at this early point in time. A reputable scientist would find it laughable. Seriously, where are the nanomachines? Stuck on a chip in a refrigerator in the basement of some university department? Funding the study of futures like this doesn't seem responsible. Biology today is much more scary. The possible societal harm of nanotechnology pales in comparison to the prospect of an avian flu epidemic, designer babies, or proliferating suitcase nukes. Even so, the funding of bioethics is currently minimal and of questionable value.

We believe the truth is more indirect. Industry is desperate not to repeat the fiascos of genetically modified organisms (GMO) in food – now banned in Europe – and to a lesser extent nuclear energy. What better way to "educate" the public then to fund a mass of public affairs and ethics professors to study the societal effects of nanotechnology? People tend to be more positive toward ideas when financially dependent on them.

Let's be clear, many of the fears disseminated to the public via science fiction and overzealous technology pundits or "futurists" are irrational; here education is important and worthwhile. Nor are professors and thinkers doing anything untoward. But it is interesting to note that the calls for all-out moratoriums on nanotechnology, ridiculous though they are, come exclusively from independent non-profit groups. So is the government's largesse enlightened or subtle propaganda? We can't know, but a prominent political theorist, Landon Winner, has called on the United States Congress not to create a nano-ethicist full employment act [Winn03]. What are important in nanotechnology are science, the progress, and the serious considerations of how to manage the environmental and social changes it creates as it develops.

When it comes down to it, saying no to nanotechnology may be like saying no to tackling the biggest problems of our time. We are faced with epic shortfalls in energy and clean water, and environmental pollution. Nanotechnology related advances in our understanding of ever-interconnected physical, material, chemical, and biological systems are our greatest hope for beating these huge problems. We must not stop science and technology.

## 1.3   What's in a name?

When asked about nanotechnology in an interview for the San Francisco Chronicle in 2004 former *Intel* CEO Craig Barrett made some illustrative remarks that are worth repeating:

> "Nanotechnology is a buzzword that you in the press have popularized, and the government popularizes it. The formal definition of nanotechnology is anything below 100 nanometers. Every transistor that we make is below 100 nanometers. So, if you want to know the investment that we're making in nanotechnology, it's the total investment that Intel is making. ... We don't happen to be the nanotechnology that you popularize with carbon nano-tubes and quantum dots and organic molecules that are going to replace CMOS transistors. Most of that is an esoteric and populist impression of what nanotechnology is. The bulk of nanotechnology is what companies like Intel and Texas Instruments and others do today."

Is Intel really the largest nanotechnology company on earth? Is everything smaller than 100 nm nanotechnology? Or is this just CEO spin? Certainly under our bipolar transition of nanotechnology above – the combination of vision and reality – Intel's latest and greatest CMOS transistor evolution is not nanotechnology. It's just the logical progression of microtechnology. First of all, companies like Intel are not taking advantage of any new nanoscale phenomena. In fact, quantum effects (like leakage via electron tunneling) are a major problem in the most advanced integrated-circuit design. Nor are they building transistors from the ground up atom-by-atom with molecules. It may be that as the biggest dimension of Intel's transistors fall below the 10 nm length scale they will have to make use of quantum size effects or switch to another paradigm – then maybe they can claim the title.

Definitions matter in contexts relevant to nanotechnology, including patenting, risk analysis, proper funding allocation, and education. Like it or not, nanotechnology is largely perceived by the public in two ways. One, as Star Trek and most science fiction conveys it: "machines you can't see" – the more fantastic vision perspective. Nanotechnology has become what we imagine it to be. From the self-cleaning walls, to the nanites of Star Trek, to the grow anything future of The Diamond Age (Stephenson, 1995). Second, the public sees the toxicology issue – dangerous particles in food and cosmetic products. This is exaggerated by the wide promotion of the term as marketing tool whether justified or not. The Woodrow Wilson Center has compiled a database of all products that claim to use nanotechnology (WWC, 2005). But the companies that are actually using the nanotechnology of today – nanoparticles – in situations that may be hazardous (such as in cosmetics) do not advertise this fact.

Nanotechnology has clearly caught the public's imagination, but they are missing the great stuff – the reality. There is a complete disconnect between what the public perceives and fears and what government agencies are defining as nanotechnology. Is the public seeing more than the toxicology or nano-robot issues? The last century taught us that chemicals could do truly awful things. Can nanotoxicity really be all that worse or is it just a scary name for more of the same? Meanwhile, governments and corporations around the world are funding nanotechnology for the promise of the next big thing and as a unifying theme in materials science development. Understanding how society – through the policy makers, thinkers, scientists, engineers, speakers – are defining and shaping the umbrella that is nanotechnology is our goal.

2. **Definitions *ad infinitum* (and more politics)**

Definitions for nanotechnology abound from government program managers and institutions, corporate entities, individual researchers, and non-profits. Let us collect and analyze some of them here. Since these are the institutions that are driving the debate and/or funding the evolution of nanotechnology, this is a worthy goal (though somewhat tedious).

The strongest shaper of the direction of nanotechnology is the United States federal government acting through its research funding agencies, primarily the National Science Foundation and various military equivalents (DARPA, ARDA, ARO, NRO, etc.). Mihail Roco, current head of the National Nanotechnology Initiative (NNI) and cheerleader for it even before its inception under President Clinton has given a number of definitions of nanotechnology that are often quoted. Since the NNI is shaping the future of nanotechnology by the power of federal grants, this is a good place to start. Roco's definitions are:

> "Nanotechnology is the creation of functional materials, devices, and systems through control of matter on the nanometer length scale, exploiting novel phenomena and properties (physical, chemical, biological) present only at that length scale." (Roco, 2000)

There is an alternate definition as well:

> "The field of nanotechnology deals with materials and systems having these key properties: they have at least one dimension of about one to 100 nanometers, they are designed through processes that exhibit fundamental control over the physical and chemical attributes of molecular-scale structures, and they can be combined to form larger structures." (SciAm, 2002)

And another:

> "Nanotechnology is the ability to understand, control, and manipulate matter at the level of individual atoms and molecules, as well as at the "supramolecular" level involving clusters of molecules. Its goal is to create materials, devices, and systems with essentially new properties and functions because of their small structure." (Roco, 2004)

According to Roco, the NNI definition encourages new contributions that were not possible before:

- "novel phenomena, properties and functions at nanoscale, which are non-scalable outside of the nanometer domain;
- the ability to measure / control / manipulate matter at the nanoscale in order to change those properties and functions;
- integration along length scales, and fields of application."

Obviously Roco is firmly in the reality camp of nanotechnology, as are most of the definitions to be described. Before we analyze it further let's look at some similar definitions. The United Kingdom, which has been ahead of the game in considering the implications and risks of nanotechnology, put together a widely cited report in 2004. The Royal Society of London report defines nanotechnology as:

> "Nanoscience is the study of phenomena and manipulation of materials at atomic, molecular, and macromolecular scales, where properties differ significantly from those at a larger scale." (Royal Society, 2004)

This definition is often quoted outside of the United States.

Roco is seeking to ignite a new "man-on-the-moon" scale investment in science and technology in this country. He is looking at nanotechnology as a unifying theme to advance all advanced technologies, which are based largely on increased understanding of the sciences made in the past century. His definition is very biased and broad to encourage these opportunities. Business as well has a motivation to make nanotechnology widely used, at least for now, because "cool" sells.

**Table 2: Four Generations of Nanotechnology according to NNI: Timeline for beginning of industrial prototyping and technology commercialization** (Roco, 2004)

$1^{st}$: Passive nanostructures. Example: coatings, nanoparticles, nanostructured metals, polymers, ceramics
$2^{nd}$: Active nanostructures. Example: transistors, amplifiers, targeted drugs, actuators, adaptive structures
$3^{rd}$: Systems of nanosystems. Example: guided assembling, 3D networking and new hierarchical architectures, robotics, evolutionary
$4^{th}$: Molecular nanosystems. Example: molecular devices 'by design', atomic design, emerging functions

**Table 3: *Nature Nanotechnology* sub-fields suitable for submission** (NN, 2005)

- Nanomaterials and nanoparticles
- Carbon nanotubes and fullerenes
- Organic-inorganic nanostructures
- Structural properties
- Electronic properties and devices
- Nanomagnetism and spintronics
- Photonic structures and devices
- Quantum information
- Molecular self-assembly
- Molecular machines and nanoelectromechanical devices (NEMS)
- Surface patterning and imaging
- Nanofluidics, nanosensors and other devices
- Nanobiotechnology and nanomedicine
- Computational nanotechnology
- Nanometrology and instrumentation
- Synthesis and processing

Recently, the elite academic publisher *Nature (London)* spun off a new journal called *Nature Nanotechnology.* It's illustrative to find out what they consider acceptable for submission. In *Nature's* words, *Nature Nanotechnology* "is a multidisciplinary journal that publishes papers of the highest quality and significance in all areas of nanoscience and nanotechnology. The journal covers research into the design, characterization and production of structures, devices and systems that involve the manipulation and control of materials and phenomena at atomic, molecular and macromolecular scales. Both bottom-up and top-down approaches - and combinations of the two – are covered." The topics included are listed in Table 3.

Meanwhile, the USA Patent and Trademark Office, trying to reorganize to deal with nanotechnology patents and simplify nanotech investment, has it's own definition:

> "Nanotechnology is related to research and technology development at the atomic, molecular or macromolecular levels, in the length of scale of approximately 1-100 nanometer range in at least one dimension; that provide a fundamental understanding of phenomena and materials at the nanoscale; and to create and use structures, devices and systems that have novel properties and functions because of their small and/or intermediate size." (USPTO, 2006)

The FDA, concerned with human exposure, takes a similar approach.

> "The FDA defines "nanotechnology" as research and technology or development of products regulated by the FDA that involve all of the following:
>
> 1) the existence of materials or products at the atomic, molecular or macromolecular levels, where at least one dimension that affects the functional behavior of the drug/device product is in the length scale range of approximately 1-100 nanometers;
> 2) the creation and use of structures, devices and systems that have novel properties and functions because of their small size; and,
> 3) the ability to control or manipulate the product on the atomic scale." (FDA, 2006)

We see a lot of similarities in all of the above definitions. We also see where Craig Barrett's confusion comes from. These definitions are so broad as to encompass virtually all developments in materials science and biotechnology. For example, although most of them make a point of requiring a "new" property to be considered, what is new?

It's not clear whether simply making something small (even in only one dimension) is enough to be considered nanotechnology. Consider two additional contrarian definitions. Two of the most vocal non-profits critical of unfettered nanotechnology investment are Greenpeace and ETC Group. Greenpeace's definition is:

> "The most common definition of nanotechnology is that of manipulation, observation and measurement at a scale of less than 100 nanometers (one nanometer is one millionth of a millimeter). However, the emergence of a multi-disciplinary field called 'nanotechnology' arises from new instrumentation only recently available, and a flow of public money into a great number of techniques and relevant academic disciplines in what has been described as an 'arms race' between governments. Nanotechnology is really a convenient label for a variety of scientific disciplines which serves as a way of getting money from Government budgets." (Huw Arnall, 2003)

The ETC Group definition:

> "Nanotechnology refers to the manipulation of living and non-living matter at the level of the nanometer (nm), one billionth of a meter. It is at this scale that quantum physics takes over from classical physics and the properties of elements change character in novel and unpredictable ways." (ETC, 2003)

The ETC Group definition is focused only on the quantum nature of some nanoscale phenomena, accounting for only a part of the reality side of nanotechnology. The Greenpeace definition is very politically astute. It's interesting though that neither include in their definition anything about risk.

It will prove interesting to check the longevity of these definitions a decade from now. They are evidence of the shotgun approach the professional community is taking toward nanotechnology. There has been some backlash to this from some of the scientific community. One example is a recent *Nature* commentary by French nanoscientist Christian Joachim. He suggests defining nano as:

> "Nanoscience should be reserved solely for the study of a single atom or a single molecule, that is, of one entity at a time, not for groups of such entities where statistics or interactions between them come into play." (Joachim, 2005)

This too narrow definition has good intentions. The ambiguity of the term nanotechnology sometimes attracts students with misperceptions about what they are getting into. Joachim suggests that a student might find themselves in a "physics of microelectronic devices" class instead of one on imagined nano-machines. Of course, one could argue that attracting students to take a physics class is good, misperception or not.

We will give a more targeted definition of what nanotechnology is today in the Conclusions of this chapter.

3.     **Perspectives from science**

Most nanotechnology definitions are cast in the language of application. This is good, since we are trying to describe a technology that does have applications now and in the future. Indeed, the language of application may be what unites all the subfields that make up nanotech and makes it a viable societal construct, though not necessarily a scientific one. Nevertheless, there are serious reasons not to lump too broad an array of advanced technologies under one banner. Education is one example. Patenting is another. Labeling inventions "nanotechnologies" sounds like a good idea for promoting interest and investment in a new and promising arena, but divorcing a patent from the specific field or principle it derives from may be devastating to the patent process. Adequately judging prior work, quality, and worthiness may depend crucially on expertise in a specific field. There exist no experts in nanotechnology, nor will there, as society has defined the term.

Here we want to comment on the "new"-ness of "new nanoscale phenomena", which is central to so many of the definitions of nanotech above. We also want to provide some context for the scientific sub-fields that nanotechnology is starting to claim, and show for a few examples where this is and is not worthwhile.

Condensed matter (CM) physics, formerly known as solid-state physics, is the largest subfield of physics, the modern six being astrophysics, atomic physics, biophysics, condensed matter or solid state physics, high energy physics, and nuclear physics. CM physics deals with the study of matter that is condensed, including the physics of metals, semiconductors, liquids, and emergent phenomena in many particle systems (such as superconductivity and the fractional quantum hall effect, atomic gases, as well as "soft" systems as in polymer physics). CM seeks to understand the non-classical behavior of everything, solid and liquid, that's cold (as compared to the sun) from the scale of a few atoms to something you can hold in your hand. So its purview is quite broad and overlaps very much with materials science and physical chemistry. In the CM community there is very little change of behavior due to the introduction of the word nanotechnology except indirectly via funding programs. Here, emphasis on applied and interdisciplinary work has become more prevalent in recent years.

CM physics has ridden (and driven) the trend in the electronics industry towards miniaturization to interesting discoveries at the nanoscale, including quantization of electrical and thermal conductance and coulomb blockade in lateral quantum dots, though often at extremely low temperatures. When CM physicists consider nanotechnology or nanoscience they usually think of one of three things: 1) quantum size effects that emerge at the nanoscale, 2) new surface physics/energetics/mechanics in nanoscale structures due to the increased surface to volume ratio, and 3) quantum coherent transport effects through mesoscale systems. Largely these occur in metallic and crystallographic systems, the most common of the latter being silicon, gallium arsenide, and carbon.

There is also a history of studying extremely small life – "nanno-biology" – but this history has been largely forgotten. The nano-bio-technology we see today in electrical engineering labs that are going "wet" is really a merger of hard inorganic device (CM) physics being applied to biological matter.

### 3.1   Cold or hot, quantum or not

One could easily fill an office with books on only one topic: silicon. Probably the most economically important and well-studied material on earth, it has taken almost 100 years (starting with the discovery of quantum mechanics) to almost fully understand this simple periodic structure. (If it has taken this long to master what's basically a simpler version of sand, how long will it take to understand biological systems?) Consider an atom of silicon. It has discrete energy levels or "shells" where electrons can exist. As the atoms come together to form a crystal, the energy levels become energy bands, continuous as a function of the electron's position within the crystal lattice. Since silicon is a semiconductor, its optical properties are largely governed by a band-gap that separates the valence (or core) electron bands from the conduction bands, where the electrons are much more mobile. Semiconductor technology is quantum, in some sense, as it requires a quantum mechanical description of matter to explain. The progression of a block of silicon crystal from the macroscopic scale down is a great way to understand some of the key manifestations of new nanoscale phenomena, but we could use many systems.

Nanotechnology proponents often bring up new quantum behavior that emerges at the nanoscale. There is nothing new, *per se*, about this phenomenon – being governed by well-known quantum physics – but only recently could matter be controlled well enough to reach this scale. There are two ways in general to see quantum effects: make a material colder or make it smaller. This is because the properties of materials, from atoms to molecules to metals and crystals, are usually governed by the interaction of electrons. And the properties of electrons are governed by the energy levels where they are allowed to exist. Crudely speaking, temperature blurs these energy levels so that electrons can hop from one level to another more easily. Here is the key point: if we make a material small enough, these energy levels will move farther apart, making this hopping due to temperature much more difficult. So, in some sense, making an object very small makes it "colder." Other new properties appear as well, such as a change in the optical excitation response – as this is defined by the spacing of these energy levels as well. This phenomena, which happens to take place around the nano-scale (say 10 nm for silicon) is what is often referred to as "quantum size effects."

## 3.2  The nano in nanoparticles

Consider nanoparticles, the nanotechnology of the present. These include nanocrystals (also called colloidal quantum dots in some cases) made of either metal or semiconductor as well as the carbon fullerene family, including buckyballs and nanotubes (the two other isotopes of carbon). Nanowires are classified under nanocrystals since they are solid objects as opposed to hollow nanotubes. All nanoparticles have at least two dimensions in the nanoscale regime, 1-100 nm. They have new properties that appear because of their size that would disappear in the micron regime. These "new properties" can be different in origin and even coincidental. We categorize them broadly into three bins: 1) new quantum effects which appear due to confinement, which are often called quantum size effects; 2) new surface physics or much increased reactivity, usually attributed to increased surface to volume ratios – this might also include structural changes from adding nanoparticles to a material (like rubber) to make it stronger; and 3) new biological properties which are largely a coincidence of scale, as the body is not attuned to defend against nanosized particles in many situations.

It would be easy to argue that nanoparticles or quantum dots are the only nanotechnology at this point in time. They certainly constitute most of the risk. Nanoparticles have large surface-to-volume ratios making them extremely reactive. They also are able to penetrate sensitive areas of the body and accumulate there. They can be embedded in larger substrates such as polymer systems or fabrics, which may decay over time and release these nanoparticles into the environment. They are also found in sunscreens and cosmetics. Nanoparticles and (eventually) active nanomachines bridge the materials to biology gap. Evolution in biomarkers for medical imaging and cancer treatment has tracked well with nanoparticle technology. But not all quantum dots are dangerous. For example, lateral quantum dots built in quantum well structures are no more toxic than the crystal they are in (gallium arsenide or silicon, as examples).

## 3.3  Not nanotechnology

Technologies are often organized by scientific principle or concept, whereas nanotech can incorporate these just because of a coincidence in scales. Take the case of photonic band gap materials, which apply the theories historically developed for electronic waves in semiconductors to photon waves (light) traveling in periodic index of refraction materials – sometimes called meta-materials. Visible light photonic band gap materials will likely require nanoscale-manufactured devices (<100 nm 3D structures) but work as well at microwave frequencies in order 1 cm sized structures. The physics doesn't change whether the features are 1 nm or 1 cm. So the only thing "nano" about some photonic meta-materials may be the fabrication process. Is there any point calling it nanotechnology? Would you send this patent application to be reviewed by the "nanotechnology" patent clerk or the physical optics patent clerk? It should be a no-brainer (the latter).

In our opinion, devices or structures with only one dimension at the nanometer length scale are in general not nanotechnology. Quantum wells are a good example. To make a quantum well you sandwich a thin semiconductor layer between two bigger semiconductor layers of a different type. This allows you to trap electrons in the inner layer. Physically this is not much different from inversion layers, where one material is grown on the other and an electric field traps electrons at the interface. These kinds of semiconductor devices are well known. Electrons in these 2D sheets exhibit many new and interesting phenomena, often attributed to the nanoscale. But, new nano phenomena in this context could just as easily be called new quantum phenomena or new many-body phenomena. Superconductivity, quantum wells and electron gasses, giant magneto-resistance (GMR) are all examples of quantum behavior misappropriated to the nano-field. This also explains some of the ridiculous money expenditure estimates – in the tens of billions of dollars some of them – which have been touted as nano-investment. If you include all the investment in GMR hard drive technology, you get big numbers.

## 4    Conclusions

Much of nanotechnology is just the natural progression of scientific and technical trends began long ago. Whether it constitutes a real paradigm shift in how we approach the manipulation and utilization of nature is still an open question. We have taken a pragmatic approach by considering the social shaping of the term as it presently stands, introducing the vision and reality viewpoints of nanotechnology. From the perspective of the scientific community, the label nanotechnology doesn't seem to be doing any harm. Funding via this route has been relatively sane, perhaps better than it would have been otherwise. Certainly investment in the cross-disciplinary developments of the fundamental science in physics, chemistry, materials science, and biology has great promise. It may even be that there is some over-investment occurring. Many states and universities across the country are building nanotechnology centers without enough qualified people to fill them.

We are left with two basic questions to ponder. How should proponents – predominately government and industry – want nanotechnology to be understood by the public? And how should we define the term in the proper sense for risk assessment? In the sole context of risk, we should immediately define a very clear definition of nanotechnology:

> **Nanotechnology, at present, is nanoparticles and nanomaterials that contain nanoparticles. Nanoparticles are defined as objects or devices with at least *two* dimensions in the nanoscale regime (typically under 10 nm) that exhibit new properties, physical, chemical, or biological, or change the properties of a bulk material, due to their size. Nanotechnology of the future will include atom-by-atom or molecule-by-molecule built active devices.**

It should be evident from the rest of our analysis that this definition is incomplete as compared to society's definition of the word. (Also note that we explicitly enforce two dimensions being in the nanoscale regime.) But it has the salience of being rigorous and relevant to the questions of nanotechnology most directly impacting the public. It might be beneficial from a public policy perspective to leave the science and technical communities alone with their developments – with their new transistors and quantum dots and so on – instead redefining nanotechnology separately, in the context only of direct and novel environmental and human impact. This would have the benefit of isolating truly worrying nanotechnology-based products from the bulk of nano-science research, which is completely innocuous. With recent calls for a moratorium on all nanotechnology by groups such as ETC, such a redefinition might have real value.

While the breadth of phenomena that lie within the nanoscale regime points to the ridiculousness of categorizing technologies based on size, the term nanotechnology has become embedded in our society, which gives it meaning explicitly. Microtechnology, as an historical counterpoint, has referred generally to a specific set of techniques and processes with little ambiguity. But what is defined as nanotechnology today may not be nanotechnology tomorrow. Taniguchi was considering precision manufacturing: separation, addition, or removal of materials at the atomic/molecular scale. Indeed, the mechanical systems that lead Taniguchi to coin the term are now called MEMS, micro-electro-mechanical systems, and are considered by many today not to be nanotechnology at all.

**Appendix: List of some technical terms**

*Spintronics:* Devices that use the electron spin instead of the electron charge for electronics. The spin is a purely quantum property of the electron with two possible states that may be used to encode information (up = 1, down = 0).

*Mesoscopics:* In physics and chemistry, the mesoscopic scale refers to the length scale at which one can reasonably discuss the properties of a material or phenomenon without having to discuss the behavior of individual atoms. For solids and liquids this is typically a few to ten nanometers, and involves averaging over a few thousand atoms or molecules. Hence, the mesoscopic scale is roughly identical to the nanoscopic or nanotechnology scale for most solids. This region often exhibits interesting quantum many-body phenomena, especially at cold temperatures.

*Quantum Computer:* A proposed computer that would exploit the quantum mechanical nature of particles, such as electrons or atomic nuclei, to manipulate information as quantum bits, or qubits. Whereas an ordinary bit has at any time a value of either 0 or 1, a qubit can also take on both values at once. Many qubits can be "entangled" (the non-local property of quantum physics) for extended interactions. Because a quantum computer can act on these multiple states simultaneously in an exponential way, it is potentially many times as powerful as a conventional computer if a proper quantum algorithm is performed.

*Quantum Well:* A quantum well is a potential well that confines particles in one dimension, forcing them to occupy a planar region, usually less than ten nanometers.

*Quantum Wire:* In material science, a quantum wire is an electrically conducting wire, in which quantum transport effects are important. Typically, this is due to the small diameter (typically of the order of nanometers) of the wires. Nanowires are a particular example of a quantum wire.

*'Lateral' Quantum Dot:* A 0-dimensional trap formed inside a semiconductor structure, the natural progression of removing another dimension of electron movement from a quantum wire. Quantum dots can exhibit properties characteristic of atoms – electron energy shells for example – and are often referred to as artificial atoms.

*Meta-material:* In electromagnetism (covering areas like optics and photonics), a meta material (or metamaterial) is an object that gains its (electromagnetic) material properties from its structure rather than inheriting them directly from the materials it is composed of. This term is particularly used when the resulting material has properties not found in naturally formed substances.

*Photonic band-gap material:* A meta-material with alternating regions of dielectric constant which can modify, trap, or guide the transport of light.

*Superconductivity:* Property of particular metals at extremely low temperatures. The electrical resistance of a conductor becomes zero, so that an electric current can flow without loss. The superfluid is an analogous quantum many-body phenomena where the friction in a liquid goes to zero.

*Transistor:* A three terminal amplifying device, the fundamental component of most active electronic circuits, including digital electronics.

*Leakage:* In the context of semiconductor electronics, leakage refers to the tunneling of electrons outside of the barriers that define a transistor for example. The result is usually heat and increased inefficiency.

*'Colloidal' quantum dot or nanocrystal:* Since the term emphasizes the quantum confinement effect it typically refers to the sub-class of nanocrystals that are small enough to exist in the quantum confinement regime, and more typically refers to fluorescent nanocrystals in the quantum confined size range.

*Nanoparticles:* Particles with controlled dimensions on the order of nanometres. Examples include colloidal gold, magnetite particles, and luminescent semiconductor aggregates that are also known as 'quantum dots'.

*Carbon fullerenes and nanotubes:* Any of various cagelike, hollow molecules composed of hexagonal and pentagonal groups of atoms, and especially those formed from carbon, that constitute the third form of carbon after diamond and graphite. Concentric shells of graphite formed by one sheet of conventional graphite rolled up into a cylinder. The lattice of carbon atoms remains continuous around the circumference. The nanotubes are designed to be very small in order to store hydrogen.

*Photonics:* The technology of transmission, control, and detection of light (photons). This is also known as fiber optics and optoelectronics.

*Electrons, photons, phonons, magnons, plasmons, polaritons, …:* Examples of fundamental excitations of nature some which are actual particles and some of which are quasi-particles (particles that represent the excitations of a sea of more fundamental particles).

*MEMS:* Tiny mechanical devices that are built onto semiconductor chips and are measured in micrometers. They are used to make pressure, temperature, chemical and vibration sensors, light reflectors and switches as well as accelerometers for airbags, vehicle control, pacemakers and games.

**Acknowledgements**

The author would like to thank Wendy Crone, Clark Miller, the other members of the Wisconsin Nanotechnology and Society Initiative, and the students of his course at the University of Wisconsin-Madison (Tahan, 2006a-b) for useful conversations. The author is supported by a USA National Science Foundation Math and Physical Sciences Distinguished International Postdoctoral Research Fellowship.

WWC (2005). Woodrow Wilson Center's Project on Emerging Nanotechnologies Consumer Product Inventory, http://www.nanotechproject.org/consumerproducts/